\documentclass[aps,prc,twocolumn,superscriptaddress,longbibliography]{revtex4-2}
\usepackage{amsmath,amssymb,physics,bm}
\usepackage{graphicx}
\usepackage{booktabs}
\usepackage{hyperref}
\usepackage{mathrsfs}
\usepackage{relsize}
\usepackage{bm}
\newcommand{\Uxi}{\Big[u(\xi)\Big]^{\raisebox{0.5pt}{\scriptsize{$n\!\!-\!\!1$}}}}

\begin{document}

\title{Newtonian I-Love-Q relations: An enthalpy-based approach}
\author{Athul Kunjipurayil and J. Piekarewicz}
\affiliation{Department of Physics, Florida State University, Tallahassee, FL 32306, USA}

\date{\today}
\begin{abstract}
 We present a pedagogical introduction to the I-Love-Q relations
 within Newtonian gravity using an enthalpy-based formulation of
 stellar structure. Introducing the specific enthalpy converts Euler's 
 equation into a first integral of hydrostatic equilibrium and recasts 
 the equation of state as a relation between the mass density and 
 the enthalpy. Together with Poisson's equation, these relations provide 
 a unified description of spherical equilibrium and deformations induced 
 by rotation and tides. We illustrate the formalism with polytropic 
 equations of state and a relativistic free Fermi gas. In Newtonian gravity, 
 slow uniform rotation and a weak tidal field excite the same linear 
 quadrupolar response, yielding the exact identity 
 $\bar I^{\,2}\bar Q/\bar\lambda\equiv1$ for the dimensionless (``bar'')
 observables---independently of the equation of state. This exact identity
 is distinguished from the approximate universality of the pairwise 
 I-Love-Q relations, whose residual dependence on stellar structure 
 is illustrated by the models considered in this work.
\end{abstract}

\maketitle

\section{Introduction}
\label{Sec:Introduction}

Among the fundamental questions highlighted in the 2015 and 2023 Long Range Plans for Nuclear Science are 
\emph{What is the nature of matter at extreme temperatures and densities?}\,\cite{LRP2015} and \emph{What 
are the nuclear processes that drive the birth, life, and death of stars?}\,\cite{LRP2023} Neutron stars---the compact 
remnants of massive stars that undergo core-collapse supernovae---provide a natural framework to address these 
questions. Despite the enormous difference in both sizes and masses, neutron stars and atomic nuclei are intimately 
connected through the strong interaction that governs the behavior of their nucleonic constituents. In neutron stars, 
quantum degeneracy and interparticle interactions generate the pressure that balances the inward pull of gravity. The 
equation of state of dense matter therefore provides the essential link between microscopic nuclear physics and 
macroscopic stellar properties. A central goal of neutron-star observations is to use this connection to constrain 
the composition of dense matter, identify signatures of possible phase transitions, and determine the equation of 
state (EOS) at densities and neutron excesses beyond the reach of terrestrial laboratories.

The microscopic underpinning of the equation of state is reflected in the rich internal structure of neutron stars. Across 
their enormous density range, matter evolves from a nonuniform crust to a uniform core, exploring many unique and
exotic phases\,\cite{Page:2006ud,Page:2009fu}. Electrons, supplemented by muons where energetically favored, 
maintain charge neutrality. This stratification provides a natural setting in which to address another central question 
in nuclear science: ``How does matter organize itself?'' From the arrangement of neutron-rich nuclei in the crust to 
the emergence of uniform matter and possible new phases, neutron stars connect the physics of finite nuclei to the 
collective behavior of dense matter\,\cite{LRP2015,LRP2023}. Relating this microscopic structure to measurable 
stellar properties is essential to extracting the underlying equation of state from astronomical observations.

Electromagnetic observations across a broad range of wavelengths have long guided and constrained theoretical models 
of neutron stars. Over the past decade, precision pulsar timing\,\cite{Cromartie:2019kug,Fonseca:2021wxt} and X-ray 
pulse-profile modeling of rotating surface hot spots\,\cite{Riley:2019yda,Miller:2019cac,Riley:2021pdl,Miller:2021qha,
Choudhury:2024xbk,Miller:2025qfq,Miller:2026vpr} have provided increasingly stringent constraints on neutron-star masses 
and radii. During this same period, gravitational-wave astronomy has provided a complementary probe of stellar structure. 
The historical detection of GW170817, the gravitational-wave signal from a binary neutron-star merger, opened a new 
observational window onto our Universe\,\cite{Abbott:PRL2017}. Of particular relevance to neutron stars is the tidal 
deformability, which characterizes the linear response of a star to the tidal field of its companion. Its strong sensitivity 
to stellar compactness allows measurements of tidal effects during the inspiral phase to constrain the neutron-star 
equation of state\,\cite{Abbott:PRL2017,Abbott:2018exr}. Understanding the dynamics underlying the equation of state 
remains a central goal of nuclear science, particularly in the poorly constrained density regime between the low-density 
domain accessible to chiral effective field theory\,\cite{Hebeler:2010jx,Tews:2012fj,Kruger:2013kua,Tews:2018kmu,
Lonardoni:2019ypg,Essick:2020flb,Drischler:2021kxf,Sammarruca:2021mhv,Sammarruca:2022ser} and perturbative 
QCD at asymptotically high densities\,\cite{Alford:1997zt,Alford:1998mk,Alford:2007xm,Annala:2019puf,
Komoltsev:2021jzg,Alford:2022bpp}. 

A complementary approach that focuses on macroscopic manifestations explores relations among stellar observables 
that are insensitive to the poorly constrained microphysics. In their pioneering work, Yagi and Yunes identified 
approximately universal relations among suitably normalized combinations of the moment of inertia, tidal deformability, 
and induced quadrupole moment---the I-Love-Q relations---that hold across a broad range of neutron-star and 
quark-star equations of state\,\cite{Yagi:2013bca,Yagi:2013awa,YAGI20171}. As quoted by the authors, these 
relations make it possible to ``learn about neutron star deformability through observations of the moment of inertia, 
break degeneracies in gravitational wave detection to measure spin in binary inspirals, distinguish neutron stars 
from quark stars, and test general relativity in a nuclear structure-independent fashion"\,\cite{Yagi:2013bca}. Their 
remarkable predictive power makes them valuable tools for connecting complementary observations; nevertheless, 
despite substantial progress in explaining their emergence, a complete physical understanding of their accuracy and 
domain of validity remains an important theoretical goal\,\cite{YAGI20171}. Motivated by these questions, we 
present a pedagogical study of the I-Love-Q relations within Newtonian gravity, using an enthalpy-based formulation 
to clarify their physical foundations and distinguish exact identities from approximately universal relations.

The individual Newtonian ingredients entering the I-Love-Q relations are well established. In particular, analytic results 
for the $n\!=\!1$ polytrope and the associated scaling relations have been already discussed in the 
literature\,\cite{Yagi:2013bca,Yagi:2013awa,Yip:2017}. Our aim is to organize these results within a unified framework 
that makes their physical connections explicit. The central tool is the specific enthalpy, which converts the Euler equation 
of hydrostatic equilibrium into a first integral and expresses the equation of state in terms of the mass density as a 
function of the specific enthalpy. Together with Poisson's equation, this formulation describes spherical equilibrium and 
its response to both slow rotations and external tidal potentials within a common set of equations.

This framework reveals a particularly simple connection between rotational and tidal deformations. In Newtonian gravity
and for a barotropic EOS, the quadrupolar part associated with slow uniform rotation has the same spatial dependence 
as an applied static quadrupolar tidal potential. Consequently, both perturbations excite the same linear quadrupolar response, 
differing only in their amplitudes. With the normalizations adopted by Yagi and Yunes, this equivalence guarantees the exact 
identity $\bar I^{\,2}\bar Q/\bar\lambda\!=\!1$, independently of the equation of state\,\cite{Yagi:2013awa,Yip:2017}. This 
exact identity must be distinguished from the approximate EOS independence of the pairwise I-Love-Q relations. In particular, 
the analytically solvable $n\!=\!1$ polytrope provides a clean demonstration of this framework and a benchmark for 
the numerical calculations. Representative polytropes and a relativistic free Fermi gas then illustrate its extension to other 
equations of state. 

The paper is organized as follows. In Sec.\ref{Sec:Formalism}, we develop an enthalpy-based formulation of Newtonian 
stellar structure built on three main ingredients: Poisson's equation, a first integral of hydrostatic equilibrium, and a barotropic 
equation of state relating the mass density to the specific enthalpy. We also show how slow uniform rotation and weak
tidal fields excite the same quadrupolar response of the star. In Sec.\ref{Sec:Results}, we present results for representative 
polytropic equations of state, making contact with the work of Yip and Leung\,\cite{Yip:2017}, and examine a relativistic Fermi 
gas whose stellar configurations interpolate smoothly between the nonrelativistic and ultrarelativistic polytropic limits. 
Finally, we summarize our findings and outline directions for future work in Sec.\ref{Sec:Conclusions}.


\section{Formalism}
\label{Sec:Formalism}

As mentioned in the Introduction, the primary goal of this paper is to develop a transparent understanding of the universal
I-Love-Q relations connecting the stellar moment of inertia $I$, the tidal Love number $k_{2}$ (or equivalently the tidal
deformability $\lambda$), and the rotation-induced quadrupole moment $Q$. Working entirely within the framework of 
Newtonian gravity, we develop a rigorous yet pedagogical framework for stellar structure that applies both to spherical 
equilibrium configurations and to any external perturbation whose associated force may be expressed as the gradient 
of a scalar potential. A central ingredient of our approach is the introduction of the specific enthalpy. For a zero-temperature 
fluid whose equation of state provides a unique relation between pressure and mass density, the enthalpy transforms the 
vector equation of hydrostatic equilibrium into a scalar first integral, thereby greatly simplifying both the formalism and its
physical interpretation.

\subsection{Newtonian Gravity}
\label{Sec:NGravity}

In the standard Newtonian formulation of stellar structure, the following are the three basic equations:
\begin{subequations}
\begin{align}
  & \nabla^{2}\Phi({\bf r})=4\pi G\rho({\bf r}), \label{HydroEqs1} \\[0.5em] 
  & \nabla P({\bf r})+\rho({\bf r})\nabla\Phi({\bf r})=0, \label{HydroEqs2} \\[0.5em] 
  & P=P(\rho) \label{HydroEqs3}.
 \end{align} 
 \label{HydroEqs}
\end{subequations}
Here $G$ is Newton's gravitational constant, $\Phi({\bf r})$ is the gravitational potential generated by the stellar mass
distribution $\rho({\bf r})$, and $P({\bf r})$ is the pressure. As such, the three equations given in Eq.(\ref{HydroEqs}) 
represent Poisson's equation, Euler's equation of hydrostatic equilibrium, and the zero-temperature equation of state, 
respectively.

In the limit of spherical symmetry---and using Gauss's law---these relations reduce to the standard formulation of Newtonian 
stellar structure:
\begin{subequations}
\begin{align}
  & \frac{dP}{dr}  = - \frac{GM(r)\rho(r)}{r^{2}}; \hspace{3pt} P(r\!=\!0)\!=\!P_{\rm c}, \label{SphSym1} \\[0.4em] 
  & \frac{dM}{dr} = 4\pi r^2\rho(r); \hspace{3pt} M(r\!=\!0)\!=\!0. 
  \label{SphSym}
 \end{align} 
\end{subequations}
Supplemented by an equation of state, this set of equations constitutes the complete Newtonian framework for stellar structure. 
The role of the equation of state is to close the system of equations by providing a relation between the pressure and the mass 
density, thereby allowing the differential equations to be integrated. The integration begins by specifying the central pressure at
the origin, as in Eq.(\ref{SphSym1}). Given that the right-hand side of both differential equations also depends explicitly on the 
mass density, the corresponding central density must then be obtained from the equation of state:
\begin{equation}
 \rho(r\!=\!0)=\rho_{\rm c}=\rho(P_{\rm c}),
\end{equation}
where we have assumed that the equation of state provides a unique relation between pressure and density. At each subsequent 
integration step, the local pressure determines the corresponding density through the relation
\begin{equation}
 \rho(r)=\rho\big(P(r)\big).
\end{equation}
This relation is applied until the stellar surface, defined by $P(r\!=\!R)\!=\!0$, is reached.

Although this formulation provides a complete description of a static, spherically symmetric star, modeling the response of the star
to either slow rotation or an external tidal field requires extending the hydrostatic equations introduced in
Eq.~(\ref{HydroEqs}).  A particularly convenient formulation is obtained by recognizing that any external perturbation generated by
a conservative force, namely, one that may be written as the gradient of a scalar potential, can be incorporated into the hydrostatic 
equations through an effective potential. The Euler equation therefore retains the same mathematical form as in Eq.(\ref{HydroEqs2}), 
provided the gravitational potential is replaced by
\begin{equation}
 \Phi_{\rm eff}({\bf r}) = \Phi({\bf r}) + \Phi_{\rm ext}({\bf r}),
 \label{PhiEff}
\end{equation}
where $\Phi({\bf r})$ denotes the self-generated gravitational potential and $\Phi_{\rm ext}({\bf r})$ represents an externally prescribed 
potential whose explicit form depends on the perturbation under consideration.

\subsection{Enthalpy and the First Integral \\ of Hydrostatic Equilibrium}
\label{Sec:Enthalpy}

By incorporating any external perturbation whose associated force may be written as the gradient of a scalar potential, the 
structure equations retain exactly the same form as those describing an isolated star. The only modification is that the 
self-generated gravitational potential is replaced by the effective potential introduced in Eq.(\ref{PhiEff}). Accordingly, the 
equation of hydrostatic equilibrium becomes
\begin{equation}
\nabla P({\bf r})+\rho({\bf r})\,\nabla\Phi_{\rm eff}({\bf r})=0.
\label{HydroEff}
\end{equation}

The key simplification arises when the pressure is uniquely determined by the mass density, namely, $P\!=\!P(\rho)$. At zero 
temperature this condition is naturally satisfied, since all thermodynamic quantities depend on a single intensive thermodynamic 
variable. It then becomes advantageous to introduce the enthalpy of the system, which  is defined as the Legendre transform of 
the internal energy. That is,
\begin{equation}
H(P,S,N)\equiv E(V,S,N)+PV,
\end{equation}
whose differential follows directly from the first law of thermodynamics:
\begin{equation}
dH=V\,dP+T\,dS+\mu_{\rm c}\,dN,
\label{Enthalpy1}
\end{equation}
where $V$ is the volume, $S$ the entropy, $N$ the number of particles, $T$ the temperature, and we denote the chemical
potential by $\mu_{\rm c}$, to distinguish it from the dimensionless mass $\mu$ introduced later.  Note that we have adopted 
the standard definition of the chemical potential,
\begin{equation}
 \mu_{\rm c}=\left(\frac{\partial E}{\partial N}\right)_{\!V,S}.
\label{ChemPot}
\end{equation}
Introducing the specific enthalpy $h=H/M$ and entropy $\sigma=S/M$, with $M=mN$ and $\rho=M/V$, one obtains
\begin{subequations}
 \begin{align}
 dh &= \frac{dP}{\rho} + T\,d\sigma \longrightarrow \frac{dP}{\rho}, \\
 h & =\frac{\mu_{\mathrm{c}}-mc^2}{m} + \sigma T \longrightarrow \frac{\mu_{\mathrm{c}}-mc^2}{m},
 \end{align}
 \label{Enthalpy2}
\end{subequations}
where $m$ is the constituent mass and the arrows indicate the zero-temperature limit. In particular, the first relation 
may be integrated to define the specific enthalpy as a function of the pressure:
\begin{equation}
 h(P) = \int_{0}^{P}\frac{dP'}{\rho(P')},
\label{EnthalpyDef}
\end{equation}
where the lower integration limit has been chosen so that the enthalpy vanishes at the stellar surface. Because the fluid is
barotropic, both the pressure and the enthalpy depend only on the mass density. Consequently, one obtains
\begin{equation}
\nabla h = \frac{dh}{dP}\,\nabla P = \frac{\nabla P}{\rho}. 
\label{GradEnthalpy}
\end{equation}
Substituting Eq.(\ref{GradEnthalpy}) into Eq.(\ref{HydroEff}) immediately yields
\begin{equation}
 \nabla\!\left(h+\Phi_{\rm eff}\right)=0 \longrightarrow h({\bf r})+\Phi_{\rm eff}({\bf r})=C,
\label{FirstIntegral}
\end{equation}
where the constant $C$ is fixed by the boundary conditions at the stellar surface. The specific enthalpy thus emerges as 
the natural thermodynamic variable for stellar structure, transforming the differential equation of hydrostatic equilibrium into 
its first integral. Accordingly, the equations of stellar structure may be written in the following equivalent form:
\begin{subequations}
\begin{align}
& \nabla^{2}\Phi({\bf r})=4\pi G\rho({\bf r}), \label{Hydro1} \\
& h({\bf r})+\Phi_{\rm eff}({\bf r})=C, \label{Hydro2} \\
& \rho=\rho(h), \label{Hydro3}
\end{align}
\label{Hydros}
\end{subequations}
where the equation of state has been recast to express the mass density as a function of the specific enthalpy. An important 
advantage of this formulation is that with the exception of Poisson's equation for the gravitational potential, all other variables
satisfy purely algebraic relations. As will become evident in the following sections, this compact formulation provides a unified 
and particularly convenient framework for describing both isolated stars and stars subject to conservative external perturbations, 
such as uniform rotation or tidal fields.


\subsection{Polytropic Equations of State}
\label{sec:Polytrope}

In this section we develop the general solution for spherical stars described by polytropic equations of state of index 
$n$ within the enthalpy formulation. Beyond its intuitive nature and simplicity, we will illustrate how the enthalpy formulation 
provides a unified and physically transparent framework for describing both the equilibrium stellar configuration and its response 
to external perturbations, particularly slow rotation and tidal perturbations. As we shall see, once the equilibrium configuration 
has been determined, both rotational and tidal deformations are governed by the same dimensionless quadrupole equation, 
thereby illustrating the close connection between seemingly distinct stellar responses.

We start by considering a polytropic equation of state of the form
\begin{equation}
 P(\rho)=K\rho^{1+1/n}\equiv K\rho^{\gamma},
\label{PolyEOS}
\end{equation}
where $K$, $n$, and $\gamma\!=\!(1\!+\!1/n)$ are the polytropic constant, the polytropic index, and the adiabatic exponent, 
respectively. For a polytropic equation of state, the integral in Eq.(\ref{EnthalpyDef}) may be evaluated in closed form and 
readily inverted to express the density directly in terms of the enthalpy. That is,
\begin{equation}
 h(\rho) = (n+1)K\rho^{1/n} \iff \rho(h)=\left[\frac{h}{(n+1)K}\right]^{n}.
\label{RHOofH}
\end{equation}

\subsection{Static Equilibrium}
\label{sec:Static}

The remarkable simplification of the enthalpy formulation now becomes apparent. In the absence of external perturbations, 
so that $\Phi_{\rm eff}\!\equiv\!\Phi$, the first integral of hydrostatic equilibrium $h+\Phi=C$, together with Poisson's equation 
for the gravitational potential, yields the following differential equation for the specific enthalpy:
\begin{equation}
 \nabla^{2}h({\bf r})=-4\pi G\rho({\bf r}) = -4\pi G\left[\frac{h({\bf r})}{(n+1)K}\right]^{n}.
 \label{DiffEq1}
\end{equation}

Assuming spherical symmetry, all thermodynamic quantities depend only on the radial coordinate $r\!=\!|{\bf r}|$, 
and Eq.(\ref{DiffEq1}) reduces to the following nonlinear second-order differential equation:
\begin{equation}
\frac{1}{r^{2}}\frac{d}{dr}\left(r^{2}\frac{dh}{dr}\right) =-\frac{4\pi G}{[(n+1)K]^{n}}h^{n}(r).
\label{DiffEq2}
\end{equation}
To expose the universal structure of this equation, we scale the enthalpy by its central value and introduce a corresponding 
dimensionless radial coordinate:
\begin{equation}
 h(r)\equiv h_c u(\xi) \hspace{5pt}{\rm and}\hspace{5pt} r\equiv a\xi,
\label{Scaling}
\end{equation}
where $h_c\!\equiv\!h(r\!=\!0)$ is the central enthalpy and the arbitrary radial scale $a$ will be chosen to simplify
the dimensionless form of the differential equation. Substitution of Eq.(\ref{Scaling}) into Eq.(\ref{DiffEq2}) yields
\begin{equation}
 \frac{1}{\xi^{2}}\frac{d}{d\xi}\left(\xi^{2}\frac{du}{d\xi}\right) =
 -\left(\frac{4\pi G a^{2}h_c^{\,n-1}}{[(n+1)K]^n}\right)\!u^{n}(\xi).
\label{DiffEqScaled}
\end{equation}
Given that the natural length scale $a$ in the problem is still at our disposal, we define it in such a way as to make 
the dimensionless combination of constants appearing on the right hand side of Eq.(\ref{DiffEqScaled}) equal to one. 
As such, the radial scale absorbs all dimensional quantities appearing in the differential equation and becomes
equal to
\begin{equation}
 a^{2}\equiv \frac{[(n+1)K]^n}{4\pi G h_c^{\,n-1}} =\frac{h_c}{4\pi G\rho_c},
\label{RadialScale}
\end{equation}
where the second equality follows directly from Eq.(\ref{RHOofH}). After these substitutions, the dimensionless 
differential equation assumes the following simple form:
\begin{equation}
 \frac{1}{\xi^{2}}\frac{d}{d\xi}\left(\xi^{2}\frac{du}{d\xi}\right)+u^{n}(\xi)=0,
\label{DimensionlessEnthalpy}
\end{equation}
with boundary conditions given by 
\begin{equation}
 u(\xi\!=\!0)=1 \hspace{5pt}{\rm and}\hspace{5pt} u'(\xi\!=\!0)=0.
\label{EnthalpyBC}
\end{equation}
Although mathematically equivalent to the classical Lane--Emden construction\,\cite{Lane1870,Emden1907}, the present 
derivation places the specific enthalpy---rather than an auxiliary dimensionless function---at the center of the analysis. As 
a result, the stellar profiles emerge directly from the fundamental equations of stellar structure. Note that all dependence 
on the dimensional quantities $K$, $G$, and $h_{c}$ (or $\rho_c$) has been absorbed into the natural scales of the problem, 
leaving a dimensionless equation whose solution depends solely on the polytropic index $n$. In particular, the stellar radius 
is identified as the first zero of the dimensionless enthalpy, namely,
\begin{equation}
  R = \zeta a  \hspace{5pt}{\rm with}\hspace{5pt} u(\zeta)=0.
\label{SurfaceXi}
\end{equation}
In turn, the total stellar mass is given by
\begin{equation}
  M = 4\pi\int_{0}^{R}\!\!\rho(r)r^{2}dr = 4\pi a^{3}\rho_{c}\int_{0}^{\zeta}\!\!u^{n}(\xi)\xi^{2}d\xi. 
\label{TotalMass1}
\end{equation}
By using the dimensionless form of the differential equation, the above integral can be evaluated in closed form: 
\begin{equation}
  \frac{M}{M_{0}} \equiv \mu = - \zeta^{2}u'(\zeta) \hspace{5pt}{\rm where}\hspace{5pt} M_{0}\!\equiv\!4\pi a^{3}\rho_{c}.
\label{TotalMass2}
\end{equation}
Thus, for a given polytropic index $n$, both the dimensionless stellar radius $\zeta$ and the dimensionless mass $\mu$
are determined entirely by the solution of Eq.(\ref{DimensionlessEnthalpy}) at the stellar surface.

\subsection{Slow Uniform Rotation}
\label{sec:Rotation}

To compute the stellar response to slow uniform rotation about the $z$ axis (${\bm\omega}\!=\!\omega\hat{\bf z}$), a new 
equilibrium configuration must be determined, as each fluid element located at ${\bf r}$ is now subjected to the centripetal 
acceleration
\begin{equation}
  {\bf a} = {\bm \omega}\!\times\!\Big({\bm \omega}\!\times\!{\bf r}\Big) = 
  -\omega^{2}s\hat{\bf s} = -\nabla\left(\frac{1}{2}\omega^{2}s^{2}\right), 
\label{aC}
\end{equation}
where $s\!=\!r\sin\theta$ is the perpendicular distance to the rotation axis and $\hat{\bf s}$ is the corresponding radial unit 
vector in the $x\!-\!y$ plane. A critical feature of the rotational acceleration is that it may be written as the gradient of a scalar 
function and may therefore be incorporated directly into the enthalpy formulation through the effective potential introduced 
in Eq.(\ref{PhiEff}). Consequently, among the three stellar structure equations listed in Eq.~(\ref{Hydros}), only the first integral 
of hydrostatic equilibrium acquires an explicit rotational contribution:
\begin{equation}
 h({\bf r})+\Phi({\bf r})-\frac{1}{2}\omega^{2}r^{2}\sin^{2}\theta=C.
\label{HRot}
\end{equation}
Unlike the static configuration, rotation introduces an explicit angular dependence into the first integral of hydrostatic equilibrium. 
However, since the rotational contribution is proportional to $\sin^{2}\theta$, its angular dependence is entirely contained in the
first two even-parity Legendre polynomials $P_0$ and $P_2$. Hence, using the identity
\begin{equation}
 \sin^{2}\theta=\frac{2}{3}\Big[1-P_{2}(\cos\theta)\Big],
\label{P2}
\end{equation}
we observe that the rotational perturbation contains only monopole ($l\!=\!0$) and quadrupole ($l\!=\!2$) components. The monopole 
contribution produces an $\mathcal{O}(\omega^{2})$ readjustment of the spherically symmetric configuration, whereas the quadrupole 
component distorts the stellar shape and generates the induced mass quadrupole moment ``Q'' that appears in the I-Love-Q relation. 
As such, the monopole sector will not be considered further.

Although Poisson's equation listed in Eq.(\ref{Hydro1}) remains unchanged, its $l\!=\!2$ component must now be solved to 
determine the induced quadrupole moment. Projecting both Poisson's equation and the first integral of hydrostatic equilibrium 
onto the $l\!=\!2$ sector yields
\begin{subequations}
\begin{align}
  \left[\frac{1}{r^2}\frac{d}{dr}\left(r^2\frac{d}{dr}\right)-\frac{6}{r^2}\right]\Phi_2(r)&= 4\pi G\rho_{2}(r), \\
  h_{2}(r)+\Phi_{2}(r)+\frac{1}{3}\omega^{2}r^{2} &= 0.
\label{L2Sector}
\end{align}
\end{subequations}

To close the system of equations, we recall that the equation of state has already been expressed in the form 
$\rho\!=\!\rho(h)$. Hence, to leading order in the rotational perturbation, a small change in the enthalpy produces 
a corresponding small change in the density:
\begin{equation}
  \rho(h+\delta h)\!=\!\rho(h)\!+\!\left(\frac{d\rho}{dh}\right)_{\!\!0}\delta h \implies
  \delta\rho\!=\!\left(\frac{d\rho}{dh}\right)_{\!\!0}\delta h,
\label{LinearEOS}
\end{equation}
where the subscript ``0'' indicates that the derivative must be evaluated along the equilibrium stellar 
configuration obtained in Sec.\ref{sec:Static}. Since this term is spherically symmetric, the different 
multipoles do not mix, and the $l\!=\!2$ components satisfy
\begin{equation}
 \rho_2(r) = \left(\frac{d\rho}{dh}\right)_{\!\!0}h_2(r) = 
 -\left(\frac{d\rho}{dh}\right)_{\!\!0}\Big[\Phi_{2}(r)+\frac{1}{3}\omega^{2}r^{2}\Big],
\label{Rho2h2}
\end{equation}
where to derive the last term we used Eq.(\ref{L2Sector}). Note that although the linear relation between 
$\rho_2$ and $h_2$ is generic, its proportionality coefficient retains its sensitivity to the underlying EOS. 
In particular, for a polytropic EOS we obtain
\begin{equation}
  \left(\frac{d\rho}{dh}\right)_{\!\!0} = \left(\frac{\rho_{c}}{h_{c}}\right)n\Uxi.
\label{DrhodH}
\end{equation}

Substituting the above two equations into Poisson's equation yields a linear, inhomogeneous ordinary 
differential equation for the rotationally induced quadrupole component of the gravitational potential. 
By reintroducing the dimensionless radial coordinate $r\!=\!a\xi$, the differential equation takes the 
following form:
\begin{equation}
 \hspace{-6pt}\left({\cal D}_{2}(\xi)\!+\!n\Uxi\right)\!\Phi_2(\xi)\!=\!-\frac{1}{3}\omega^{2}a^{2}n\Uxi\xi^{2},
\label{Phi2Equation}
\end{equation}
where we have defined the $l\!=\!2$ component of the dimensionless Laplacian operator as
\begin{equation}
 {\cal D}_{2}(\xi)\!\equiv\!\left[\frac{1}{\xi^{2}}\frac{d}{d\xi}\left(\xi^{2}\frac{d}{d\xi}\right) -\frac{6}{\xi^{2}}\right].
\label{Laplace2}
\end{equation}

Finally, we define the dimensionless quadrupole component of the gravitational potential by factoring out the 
natural strength of the rotational perturbation:
\begin{equation}
 \Phi_2(\xi)\equiv\frac{1}{3}\omega^{2}a^{2}\varphi(\xi).
\label{SmallPhi2}
\end{equation}
This definition reduces Eq.(\ref{Phi2Equation}) to the fully dimensionless form
\begin{equation}
 \left({\cal D}_{2}(\xi)+n\Uxi\right)\varphi(\xi) =-n\Uxi\xi^{2}.
\label{ScaledPhi2Equation}
\end{equation}
The solution to this inhomogeneous differential equation may be written as the sum of a homogeneous and 
a particular solution. Denoting the regular homogeneous solution by $\psi(\xi)$, the general interior solution 
takes the form
\begin{equation}
 \varphi_{\rm int}(\xi) =A\psi(\xi)+\varphi_p(\xi)=A\psi(\xi)-\xi^{2},
\label{Phi2InteriorScaled}
\end{equation}
where $A$ is a normalization constant. The simplicity of the particular solution follows from the fact that $\xi^2$ 
is the regular solution of Laplace's equation in the $l\!=\!2$ sector, so that
\begin{equation}
 {\cal D}_{2}(\xi)\xi^{2}=0.
\label{D2Xi2}
\end{equation}
Hence, $\varphi_p(\xi)\!=\!-\xi^2$ satisfies the full inhomogeneous equation for an arbitrary polytropic index $n$. 

Near the origin, the $1/\xi^2$ term in the Laplacian dominates, so regularity at the origin demands that the homogeneous 
solution behaves as
\begin{equation}
 \lim_{\xi\to0}\frac{\psi(\xi)}{\xi^2}=1 \hspace{5pt} {\rm or}\hspace{5pt} \psi(\xi)\underset{\xi\to0}{\sim}\xi^2.
\label{psizero}
\end{equation}

Outside the star, where the mass density vanishes, the scaled gravitational potential satisfies Laplace's equation. By
requiring that the quadrupole component of the solution vanishes at spatial infinity, one selects the decaying $l\!=\!2$ 
solution,
\begin{equation}
 \varphi_{\rm ext}(\xi)=\frac{B}{\xi^3},
\label{Phi2ExteriorScaled}
\end{equation}
where the constants $A$ and $B$ are determined by matching the interior and exterior solutions, together with their first 
derivatives, at the stellar surface $\xi\!=\!\zeta$. This yields
\begin{equation}
  A\psi(\zeta) = \left(\frac{5}{3+\eta}\right)\zeta^{2} 
  \hspace{5pt} {\rm and}\hspace{5pt} 
  B = \left(\frac{2-\eta}{3+\eta}\right)\zeta^{5},
\label{AandB}
\end{equation}
where $\eta$ is the dimensionless logarithmic derivative of the homogeneous solution evaluated at the stellar surface,
\begin{equation}
\eta \equiv \zeta\frac{\psi'(\zeta)}{\psi(\zeta)}.
\label{LogDer}
\end{equation}
Thus, all the information required to determine the rotationally induced mass quadrupole is encoded in the behavior of 
the homogeneous solution at the stellar surface.

\subsection{Induced Quadrupole Moment}
\label{sec:Quadrupole}

Finally, the induced quadrupole moment may be extracted directly from the asymptotic behavior of the dimensional 
quadrupole potential $\Phi_2$ outside the star. Using the scaling introduced in Eq.(\ref{SmallPhi2}), one finds
\begin{equation}
\Big[\Phi_2(\xi)\Big]_{\rm ext} = \frac{1}{3}\omega^{2}a^{2}\varphi_{\rm ext}(\xi) = \frac{1}{3}\omega^{2}a^{2}\frac{B}{\xi^{3}}.
\label{BigPhi2}
\end{equation}
Restoring units by using $\xi\!=\!r/a$, $\zeta\!=\!R/a$, together with Eq.(\ref{AandB}), the exterior potential may be expressed 
in terms of physical quantities as
\begin{equation} 
 \Big[\Phi_2(r)\Big]_{\rm ext} =\frac{\omega^{2}R^{5}}{3} \left(\frac{2-\eta}{3+\eta}\right)\frac{1}{r^{3}}\equiv -\frac{GQ}{r^{3}}.
 \label{BigPhi2Ext}
\end{equation}
Comparison with the standard asymptotic form of the gravitational quadrupole potential given above yields 
\begin{equation}
 Q= -\frac{\omega^{2}R^{5}}{3G}
\left(\frac{2-\eta}{3+\eta}\right).
\label{InducedQ}
\end{equation}
The negative sign reflects the oblate deformation induced by uniform rotation. Aside from the overall dimensional scale 
$\omega^{2}R^{5}/G$, the entire rotational quadrupole response is controlled by $\eta$---the dimensionless logarithmic 
derivative evaluated at the stellar surface.

\subsection{Tidal Distortions}
\label{sec:Tides}

Having developed the moment of inertia and the rotationally induced quadrupole moment in the Newtonian framework, 
we now turn to the tidal response of the star. As in the case of rotation, a major advantage of the enthalpy formulation 
is that tidal effects may be incorporated without introducing any new ingredients. The essential task is simply to identify 
the external contribution to the effective potential appearing in Eq.(\ref{Hydro2}).

To do so, we consider a star of radius $R$ placed in the gravitational field generated by a companion of mass ${\mathcal M}$ 
located at ${\bf D}\!=\!D\hat{\bf z}$ relative to the center of the star, with $D\!>\!R$. The external gravitational potential 
evaluated at a fluid element located at ${\bf r}$ relative to the stellar center is
\begin{equation}
 \Phi_{\rm ext}({\bf r}) = -\frac{G{\mathcal M}}{|{\bf D}-{\bf r}|} = 
  -G{\cal M}\sum_{l=0}^{\infty}\frac{r^{l}}{D^{l+1}}P_{l}(\cos\theta).
\label{PhiExt}
\end{equation}
The $l\!=\!0$ term is spatially constant and therefore simply redefines the zero of the gravitational potential. In turn, 
the $l\!=\!1$ term is proportional to $z$ and generates a uniform acceleration of the star as a whole. It therefore 
disappears in a reference frame centered on the star and produces no internal deformation. The leading contribution 
to the deformation of the star is thus the quadrupole ($l\!=\!2$) term:
\begin{equation}
 \Phi_{\rm tidal}({\bf r}) = -\frac{G{\mathcal M}}{D^{3}}r^{2}P_{2}(\cos\theta) 
 \equiv \frac{1}{2}{\mathcal E}r^{2}P_{2}(\cos\theta),
 \label{TidalTensor}
\end{equation}
where we have introduced the scalar tidal strength
\begin{equation}
{\mathcal E} \equiv -\frac{2G{\mathcal M}}{D^3}.
\label{TidalStrength}
\end{equation}
The tidal interaction may therefore be incorporated into precisely the same effective-potential formalism 
used for uniform rotation. The first integral of hydrostatic equilibrium now becomes
\begin{equation}
h({\bf r}) + \Phi_{\rm eff}({\bf r}) = h({\bf r}) + \Phi({\bf r}) + 
\frac{1}{2}\mathcal{E}r^2P_2(\cos\theta) = C.
\label{HTidal2}
\end{equation}
Unlike uniform rotation, whose angular dependence contains both monopole and quadrupole components, 
the leading tidal perturbation is purely quadrupolar. Thus, after removing the physically irrelevant monopole 
and dipole contributions, the companion acts on the internal structure of the star through a pure quadrupolar 
external potential. 

Following the same steps as in the previous section, we obtain a linear, inhomogeneous ordinary differential 
equation for the quadrupole component of the gravitational potential:
\begin{equation}
 \hspace{-5pt}\left({\cal D}_{2}(\xi)\!+\!n\Uxi\right)\!\Phi_2(\xi)\!=\!-\frac{1}{2}{\mathcal E}a^{2}n\Uxi\xi^{2}.
\label{EquationPhi2}
\end{equation}
Relative to Eq.(\ref{Phi2Equation}) that encodes uniform slow rotation, the only modification in the case of tidal
distortions is the replacement 
\begin{equation}
   \frac{1}{3}\omega^{2} \longrightarrow \frac{1}{2}{\mathcal E}.
\label{EOmega}
\end{equation}
This correspondence uncovers a unique feature of Newtonian gravity: the quadrupolar part associated with 
slow uniform rotation has the same spatial dependence as a static quadrupolar tidal potential, so both excite 
the same quadrupolar response, differing only in the amplitude of the perturbation. Although the above derivation 
was limited to polytropes, this correspondence holds for any barotropic equation of state. 


\subsection{Tidal Deformability}
\label{sec:k2}

Having established that the tidal response follows directly from the rotational problem through the replacement 
in Eq.(\ref{EOmega}), the tidally induced quadrupole moment may be obtained immediately from Eq.(\ref{InducedQ}):
\begin{equation}
 Q_{\rm tidal} =-\frac{{\mathcal E}R^{5}}{2G} \left(\frac{2-\eta}{3+\eta}\right).
\label{TidalQ}
\end{equation}
The applied tidal field induces a stellar mass quadrupole that in the linear regime satisfies
\begin{equation}
 Q_{\rm tidal}=-\lambda_{\rm s}\mathcal{E},
 \label{TidalLambda}
\end{equation}
where $\lambda_{\rm s}$ is the scalar tidal deformability. Comparison with Eq.(\ref{TidalQ}) gives
\begin{equation}
 \lambda_{\rm s}=\frac{R^5}{2G}
 \left(\frac{2-\eta}{3+\eta}\right).
 \label{ScalarLambda}
\end{equation}
Equivalently, the tidal response can be expressed in terms of the dimensionless quadrupolar Love number $k_2$, 
defined in the present convention by
\begin{equation}
 \lambda_{\rm s}\equiv k_2\frac{R^5}{G},
 \label{LoveDefinition}
\end{equation}
so that
\begin{equation}
 k_2=\frac{1}{2}\left(\frac{2-\eta}{3+\eta}\right).
 \label{LoveNumber}
\end{equation}
Hence, the Love number is therefore determined by the same surface quantity $\eta$ that controls the rotationally induced 
quadrupole moment, reflecting the common origin of the quadrupolar response to rotation and tides in Newtonian gravity.

Given that the scalar tidal strength introduced in Eq.(\ref{TidalStrength}) differs by a normalization from the tidal tensor 
commonly employed in the literature, the corresponding tidal deformability is also convention dependent. Adopting the 
standard tensor definition $Q_{ij}=-\lambda\mathcal{E}_{ij}$, 
the dimensional tidal deformability is related to the Love number by\,\cite{Hinderer:2007mb,Damour:2009vw,PoissonWill:2014}
\begin{equation}
 \lambda=\frac{2}{3}k^{2}\frac{R^{5}}{G} =\frac{R^{5}}{3G}\left(\frac{2-\eta}{3+\eta}\right).
\label{lambdaTensor}
\end{equation}
In the construction of the I-Love-Q relations, the associated dimensionless tidal deformability, together with the dimensionless 
moment of inertia and induced quadrupole moment, will be introduced in the next section.

\subsection{I-Love-Q Relations}
\label{sec:ILoveQ}

\subsubsection{Dimensionless Moment of Inertia}
\label{sec:DimI}

The first stellar property relevant to the I-Love-Q relations is the moment of inertia.  In Newtonian gravity it is completely 
determined by the equilibrium density profile and may therefore be computed directly from the spherical configuration 
obtained in Sec.\ref{sec:Static}:
\begin{equation}
 I = \frac{8\pi}{3}\int_{0}^{R}r^{4}\rho(r)dr = \frac{8\pi}{3}a^{5}\rho_{c}\int_{0}^{\zeta}\!\!u^{n}(\xi)\xi^{4}d\xi. 
 \label{MomI}
\end{equation}
Scaling the moment of inertia by its natural dimensional scale $MR^{2}$, we define the dimensionless structure coefficient as
\begin{equation}
  \mathlarger{\kappa}_{I}(n) \equiv \frac{I}{MR^{2}}
  = \frac{2}{3\mu\zeta^{2}} \int_{0}^{\zeta}\!u^{n}(\xi)\xi^{4}d\xi .
 \label{ScaledI}
\end{equation}
Thus $\kappa_{I}(n)$ is determined entirely by the dimensionless equilibrium solution $u(\xi)$, which itself only depends on the 
chosen polytropic index. This simplicity is unique to Newtonian gravity and stands in contrast to the general relativistic case, 
where the moment of inertia is obtained from the equations describing slow rotation\,\cite{Hartle:1967he}.

The coefficient ${\kappa}_{I}(n)$ isolates the dependence of the moment of inertia on the internal stellar structure, but has 
yet to be connected to the standard definition of the dimensionless moment of inertia used in the I-Love-Q relations. Introducing 
the stellar compactness,
\begin{equation}
 {\cal C}\equiv \frac{GM}{Rc^{2}},
\label{Compactness}
\end{equation}
the standard dimensionless moment of inertia employed in the I-Love-Q literature\,\cite{Yagi:2013bca,Yagi:2013awa} may be 
written as
\begin{align}
\bar{I} & \equiv \left[\frac{c^{4}}{G^{2}}\right] \left(\frac{I}{M^{3}}\right) \nonumber \\
           & = \mathlarger{\kappa}_{I}(n)\left[\frac{Rc^{2}}{GM}\right]^{2} \nonumber \\
           & = \mathlarger{\kappa}_{I}(n)\,{\cal C}^{-2}.
 \label{Ibar}
\end{align}
Here and throughout this section, quantities in parentheses denote the corresponding expressions in geometrized units $(G\!=\!c\!=\!1)$, 
whereas those enclosed in square brackets restore the appropriate powers of $G$ and $c$ required to write the expression in physical 
units. Note that the stellar compactness fixes the overall scaling of $\bar{I}$, whereas any residual dependence on the equation of state 
is entirely encoded in the dimensionless structure coefficient $\kappa_I(n)$.

\subsubsection{Dimensionless Induced Quadrupole Moment}
\label{sec:DimQ}

Continuing to follow the conventions of Refs.\cite{Yagi:2013bca,Yagi:2013awa}, we define the scaled induced quadrupole moment as
\begin{equation}
 \bar{Q} \equiv - \left[\frac{c^{4}}{G^{2}}\right]\left(\frac{Q}{M^{3}{\chi^{2}}}\right),
\label{Qbar0}
\end{equation}
where the dimensionless spin is defined as 
\begin{equation}
 \chi \equiv \left[\frac{c}{G}\right]\left(\frac{J}{M^{2}}\right) =  \left[\frac{c}{G}\right]\left(\frac{I\omega}{M^{2}}\right). 
\label{Spin}
\end{equation}
Using Eq.(\ref{InducedQ}) for the induced quadrupole moment and Eq.(\ref{ScaledI}) for the moment of inertia,
the properly scaled induced quadrupole moment takes the following form:
\begin{align}
  \bar{Q} & = [c^{2}]\left(\frac{M}{I^{2}}\right)\left(-\frac{Q}{\omega^{2}}\right) \nonumber \\
              & = \frac{1}{3\kappa_{I}^{2}(n)}\left(\frac{2-\eta}{3+\eta}\right)\left[\frac{Rc^{2}}{GM}\right] \nonumber \\
  	      & = \frac{1}{3\kappa_{I}^{2}(n)}\left(\frac{2-\eta}{3+\eta}\right){\cal C}^{-1} \equiv \kappa_{Q}(n)\,{\cal C}^{-1},
\label{Qbar1}
\end{align}
where we have introduced the dimensionless structure coefficient associated with the induced quadrupole moment as
\begin{equation}
  \mathlarger{\kappa}_{Q}(n) = \frac{1}{3\kappa_{I}^{2}(n)}\left(\frac{2-\eta}{3+\eta}\right).
 \label{ScaledQ}
\end{equation}
As in the case of the dimensionless moment of inertia, the compactness determines the overall scaling of $\bar Q$, 
whereas any remaining dependence on the internal stellar structure is contained in the dimensionless coefficient 
$\kappa_Q(n)$. In particular, the explicit dependence on the angular velocity has disappeared, as required in the 
slow-rotation limit, leaving a quantity that depends only on the intrinsic stellar properties. 

\subsubsection{Dimensionless Tidal Deformability}
\label{sec}

The last member of the I-Love-Q triad is the dimensionless tidal deformability defined as follows\,\cite{Yagi:2013bca,Yagi:2013awa}:
\begin{equation}
\Lambda \equiv \bar{\lambda} = \left[\frac{c^{10}}{G^{4}}\right] \left(\frac{\lambda}{M^{5}}\right).
\label{Lbar0}
\end{equation}
Using Eq.(\ref{lambdaTensor}) for the tidal deformability, we obtain the dimensionless tidal deformability entering the 
I-Love-Q relations as
\begin{equation}
\Lambda = \bar{\lambda} = \frac{1}{3}\left(\frac{2-\eta}{3+\eta}\right)\left[\frac{Rc^{2}}{GM}\right]^{5}
	      = \mathlarger{\kappa}_{\lambda}(n)\,{\cal C}^{-5},
\label{Lbar1}
\end{equation}
where the dimensionless structure coefficient associated with tidal distortions is now given by
\begin{equation}
 \mathlarger{\kappa}_{\lambda}(n) \equiv \frac{1}{3} \left(\frac{2-\eta}{3+\eta}\right).
\label{ScaledL}
\end{equation}
Thus, the compactness parameter once again determines the overall scaling, while the dependence on the internal stellar structure is 
entirely contained in the dimensionless coefficient $\kappa_{\lambda}(n)$.

Before moving to the Results section, we conclude this long formalism section by collecting all three dimensionless parameters
entering the I-Love-Q relations. From Eqs.(\ref{ScaledI}), (\ref{ScaledL}), and (\ref{ScaledQ}), we obtain
\begin{subequations}
\begin{align}
   \bar{I}  & =  \mathlarger{\kappa}_{I}(n)\,{\cal C}^{-2}, &
                      \mathlarger{\kappa}_{I}(n) & = \frac{2}{3\mu\zeta^{2}} \int_{0}^{\zeta}\!u^{n}(\xi)\xi^{4}d\xi, \\
  \bar{\lambda} &= \mathlarger{\kappa}_{\lambda}(n)\,{\cal C}^{-5},  &
                              \mathlarger{\kappa}_{\lambda}(n)&=\frac{1}{3} \left(\frac{2-\eta}{3+\eta}\right), \\
 \bar{Q} & =  \mathlarger{\kappa}_{Q}(n)\,{\cal C}^{-1}, &
                     \mathlarger{\kappa}_{Q}(n) & = \frac{1}{3\kappa_{I}^{2}(n)}\left(\frac{2-\eta}{3+\eta}\right).
\end{align}
\label{ScaledILQ}
\end{subequations}
In particular, note that the dimensionless structure coefficients are not all independent but rather satisfy the relation
\begin{equation}
 \mathlarger{\kappa}_{\lambda}(n) = \mathlarger{\kappa}_{I}^{2}(n)\mathlarger{\kappa}_{Q}(n).
\label{KappaILQ}
\end{equation}

This relation involving the three scaled structure coefficients in Eqs.(\ref{ScaledILQ}) naturally raises the following question: 
does this shared mathematical structure also explain the I-Love-Q relations?  The existence of universal relations suggests 
that, once the dominant dependence on the compactness parameter is removed, the remaining structure may be effectively 
governed by a small number of macroscopic parameters. Here we explore the possibility that the origin of this universality is 
already manifest in Newtonian gravity, where the mathematics is simpler and the physical origin may be easier to isolate. 
Indeed, Yagi and Yunes\,\cite{Yagi:2013bca,Yagi:2013awa,YAGI20171} demonstrated that these observables exhibit remarkable 
equation-of-state independence when expressed in dimensionless form. Using Eqs.(\ref{ScaledILQ}) one may then construct 
the two independent ratios
\begin{subequations}
\begin{align}
  & {\cal R}_{I\lambda}(n) \equiv \frac{\bar{I}}{\bar{\lambda}^{2/5}} = 
      \frac{\kappa_{I}}{\kappa_{\lambda}^{2/5}}, \\
  & {\cal R}_{Q\lambda}(n) \equiv \frac{\bar{Q}}{\bar{\lambda}^{1/5}} = 
      \frac{\kappa_{Q}}{\kappa_{\lambda}^{1/5}}.  
\end{align}
\label{RatiosKappa}
\end{subequations}
These relations become independent of the compactness parameter and are expected to be weakly dependent on the underlying 
equation of state. As such, they could provide the Newtonian foundation for the approximate equation-of-state independence of the relativistic I-Love-Q relations uncovered by Yagi and Yunes\,\cite{Yagi:2013bca,Yagi:2013awa}. More generally, these results may
suggest that the origin of the universality lies in the similarity of stellar density profiles once the dominant dimensional scaling with compactness has been removed. From this perspective, general relativity may be viewed not necessarily as the origin of the 
I-Love-Q relations, but rather as a refinement of a universality that is already present in the Newtonian limit. 

However, the additional identity between the dimensionless structure coefficients in Newtonian gravity leads to an even stronger 
result. Indeed, the three dimensionless observables combine to form the exact identity
\begin{equation}
  \frac{\bar{I}^{2}\bar{Q}}{\bar{\lambda}} \equiv 1,
\label{ExactILQ}
\end{equation}
first obtained in the Newtonian analysis of Yagi and Yunes\,\cite{Yagi:2013awa} and subsequently discussed in the context of 
Newtonian I-Love-Q relations in Ref.\cite{Yip:2017}. In the present formulation, the origin of this identity becomes particularly
transparent: rotational and tidal perturbations excite the same $l\!=\!2$ response, differing only in the strength of the applied 
quadrupolar potential; see Eq.(\ref{EOmega}). The dependence on the internal structure is therefore identical in both responses 
and cancels in the combination appearing in Eq.(\ref{ExactILQ}). Unlike the familiar pairwise I-Love-Q relations, whose universality 
is robust but approximate, Eq.(\ref{ExactILQ}) is an exact identity within the Newtonian framework considered here. As we show 
in the next section, this identity is not restricted to polytropic stars but holds for an arbitrary barotropic equation of state.

\section{Results}
\label{Sec:Results}

We start this section by isolating the analytically solvable $n\!=\!1$ polytrope. We then proceed to solve numerically 
the relevant equations to explore the behavior of the I-Love-Q relations for a broad range of polytropic equations of 
state. We conclude this section by demonstrating that the analytic result obtained in the previous section extends to
a relativistic Fermi gas equation of state.

\subsection{Analytic Solution for the {\boldmath $n=1$} Polytrope}
\label{sec:Analytic}

Having developed the entire formalism in the previous section, we illustrate the power of the enthalpy formulation 
by focusing on the analytically solvable $n\!=\!1\,(\gamma\!=\!2)$ polytrope. We derive in a physically transparent 
framework closed-form expressions for all the stellar observables entering the I-Love-Q relations---including the 
equilibrium structure, the moment of inertia, the rotationally induced quadrupole moment, and the tidal Love number. 
Focusing on the analytically solvable $n\!=\!1$ case, the equation of state may be recast in the form
\begin{equation}
 P(\rho)\!=\!K\rho^{2} \iff \rho(h) = \frac{h}{2K}.
\label{RhoH}
\end{equation}

\subsubsection{Static Equilibrium}
\label{sec:StaticN1}

In the absence of external perturbations, it was shown in Sec.\ref{sec:Static} that, in static equilibrium and for 
the particular case of $n\!=\!1$, the scaled enthalpy satisfies the following linear, second-order ordinary differential 
equation:
\begin{subequations}
\begin{align}
 & \frac{1}{\xi^{2}}\frac{d}{d\xi}\left(\xi^{2}\frac{du}{d\xi}\right)+u(\xi)=0, \\
 & u(\xi\!=\!0)=1 \hspace{5pt}{\rm and}\hspace{5pt} u'(\xi\!=\!0)=0,
\label{uOne0}
\end{align}
\end{subequations}
which has the analytic solution
\begin{equation}
  u(\xi)=\frac{\sin{\xi}}{\xi}.
\label{uOne1}
\end{equation}
From this solution one can obtain all relevant thermodynamic quantities:
\begin{subequations}
 \begin{align}
  h(r) & = h_{c}\left(\frac{\sin{\xi}}{\xi}\right), \\
  \rho(r) &= \rho_{c}\left(\frac{\sin{\xi}}{\xi}\right),\\
  P(r) & = P_{c} \left(\frac{\sin{\xi}}{\xi}\right)^{\!2}, \\
  M(r) & = 4\pi a^{3}\rho_{c}\Big(\sin(\xi)-\xi\cos(\xi)\Big).
 \end{align}
 \label{Profiles}
\end{subequations}
Here $h_{c}\!=\!2K\rho_{c}$, $P_{c}\!=\!K\rho_{c}^{2}$, and we have restored the physical radial coordinate 
$r\!=\!a\xi$, where the radial scale $a$ for $n\!=\!1$ is given by
\begin{equation}
  a = \sqrt{\frac{h_{c}}{4\pi G\rho_{c}}} = \sqrt{\frac{K}{2\pi G}}\,.
  \label{sqrtA}
\end{equation}
The stellar surface is located at the first zero of $u(\xi)$, namely at $\zeta\!=\!\pi$, where both the pressure and the density 
vanish. Moreover, for $n\!=\!1$ the dimensionless mass also takes the value of $\mu\!=\!\pi$. Thus, the radius and total mass 
of the star are given from Eqs.(\ref{SurfaceXi}) and (\ref{TotalMass2}) to be
\begin{equation}
  R  = \pi a \hspace{5pt}{\rm and}\hspace{5pt} M  = 4\pi^{2}a^{3}\rho_{c}.
 \label{RandM}
\end{equation}

Having obtained the closed-form solution $u(\xi)$, together with $\zeta\!=\!\mu\!=\!\pi$, we now evaluate analytically the 
dimensionless structure coefficient associated with the moment of inertia:
\begin{equation}
  \mathlarger{\kappa}_{I}(n\!=\!1) = \frac{2}{3\pi^{3}} \int_{0}^{\pi}\!\xi^{3}\sin\xi\,d\xi 
                                                    = \frac{2}{3}\left(\frac{\pi^{2}-6}{\pi^{2}}\right) \approx 0.261382.
 \label{KappaIN1}
\end{equation}

To compute the remaining two dimensionless structure coefficients, we require the logarithmic derivative of the regular solution
to the homogeneous equation in Eq.(\ref{ScaledPhi2Equation}). For $n\!=\!1$, this equation reduces to
\begin{equation}
 \Big({\cal D}_{2}(\xi)+1\Big)\psi(\xi) = 0.
 \label{ScaledPsiEquationN1}
\end{equation}
The above differential equation is the spherical Bessel equation of order $l\!=\!2$, whose regular solution is proportional to
\begin{equation}
  j_{2}(\xi) = \frac{(3-\xi^{2})\sin\xi-3\xi\cos\xi}{\xi^{3}}.
 \label{SphericalBessel2}
\end{equation}
Since $j_{2}(\xi)\sim\xi^{2}/15$ near the origin, the normalization adopted in Eq.(\ref{psizero}) corresponds to $\psi(\xi)=15j_{2}(\xi)$. 
This overall normalization, however, is irrelevant for the dimensionless structure coefficients, which depend only on the logarithmic 
derivative. At the stellar surface $\zeta=\pi$, one obtains
\begin{equation}
 \eta = \pi\frac{\psi'(\pi)}{\psi(\pi)} = \frac{\pi^{2}-9}{3}.
 \label{etaN1}
\end{equation}
Using Eqs.(\ref{ScaledILQ}), the remaining two dimensionless structure coefficients are therefore
\begin{subequations}
\begin{align}
  \mathlarger{\kappa}_{\lambda}(n\!=\!1) & = \left(\frac{15-\pi^{2}}{3\pi^{2}}\right) \approx 0.173273. \\
   \mathlarger{\kappa}_{Q}(n\!=\!1) & = \frac{3\pi^{2}(15-\pi^{2})}{4(\pi^{2}-6)^{2}} \approx 2.53617.
\end{align}
\label{ScaledKappa}
\end{subequations}
As a useful consistency check, these analytic expressions satisfy---as they must---the Newtonian identity listed in
Eq.(\ref{KappaILQ}). 

\subsection{Representative Polytropic Models}

Rather than undertaking a comprehensive survey of polytropic equations of state, which has already been explored in considerable 
detail by Yip and Leung\,\cite{Yip:2017}, we consider---in addition to the analytic $n\!=\!1$ case---two representative models: 
$n\!=\!3/2$ and $n\!=\!3$. These cases correspond to the nonrelativistic and ultrarelativistic limits, respectively, of a zero-temperature 
ideal Fermi gas. 

Table\,\ref{Table1} summarizes the properties of the static stellar configurations and their quadrupolar responses for the three polytropic equations of state. Here, $\zeta$ and $\mu$ denote the dimensionless stellar radius and mass, respectively; $\eta$ is the dimensionless logarithmic derivative of the homogeneous quadrupolar response evaluated at the stellar surface; and $k_2$ is the Love number. The dimensionless I-Love-Q structure coefficients are denoted by $\kappa_I$, $\kappa_\lambda$, and $\kappa_Q$. The final column gives the compactness-independent coefficient
\begin{equation}
 \mathcal{R}_{I\lambda}\equiv\frac{\kappa_I}{\kappa_\lambda^{2/5}},
 \label{RILambda}
\end{equation}
which enters the I-Love relation $\bar I=\mathcal{R}_{I\lambda}\bar\lambda^{2/5}$. The $n\!=\!1$ model serves as an analytic benchmark, while the $n\!=\!3/2$ model represents a moderately soft equation of state and the $n\!=\!3$ model illustrates the breakdown of universality for sufficiently soft polytropes\,\cite{Yip:2017}.

\begin{table*}[t]
\squeezetable
\begin{ruledtabular}
\begin{tabular}{ccccccccc}
& \multicolumn{2}{c}{Static spherical}
& \multicolumn{2}{c}{Quadrupolar response}
& \multicolumn{3}{c}{I--Love--Q coefficients}
& \multicolumn{1}{c}{Universality} \\
\cline{2-3}\cline{4-5}\cline{6-8}\cline{9-9}
$n$ & $\zeta$ & $\mu$ & $\eta$ & $k_2$
& $\kappa_I$ & $\kappa_\lambda$ & $\kappa_Q$ & $\mathcal{R}_{I\lambda}$ \\
\hline
$1$   & 3.14159 & 3.14159 & 0.289868  & 0.259909  & 0.261382  & 0.173273   & 2.53617 & 0.526967 \\
$3/2$ & 3.65375 & 2.71406 & 0.886340 & 0.143279  & 0.204600  & 0.0955192  & 2.28181 & 0.523443 \\
$3$   & 6.89685 & 2.01824 & 1.85963    & 0.0144430 & 0.0753576 & 0.00962865 & 1.69555 & 0.482726 \\
\end{tabular}
\end{ruledtabular}
\caption{Dimensionless structure coefficients for three representative polytropes. See the text for details.}
\label{Table1}
\end{table*}

The results in Table\,\ref{Table1} provide a clear illustration of both the success and the limitations of Newtonian I-Love-Q 
universality. The coefficient $\mathcal{R}_{I\lambda}$ changes by less than one percent between the $n\!=\!1$ and $n\!=\!3/2$ 
polytropes, whereas the $n\!=\!1$ value is about $10\%$ larger than the corresponding result for $n\!=\!3$. This behavior is 
consistent with the analysis of Yip and Leung, who showed that the I-Love-Q relations are highly insensitive to the equation 
of state for stiff polytropes, whereas universality progressively deteriorates for sufficiently soft and centrally condensed 
configurations\,\cite{Yip:2017}. In particular, the $n\!=\!3$ ($\gamma\!=\!4/3$) polytrope lies at the boundary of Newtonian 
radial stability and should be viewed as a limiting case rather than as a representative neutron-star model. 

By contrast, general-relativistic neutron-star calculations exhibit approximate I-Love-Q universality across a broad set of 
physically motivated equations of state\,\cite{Yagi:2013awa}. Thus, the present results distinguish the nearly universal 
behavior of stiff and moderately soft Newtonian stars from the expected breakdown for soft polytropes\,\cite{Yip:2017}.

\subsection{Beyond Polytropes: A Relativistic Fermi Gas}

Having explored the I-Love-Q relations for a representative set of polytropic models, we now consider a qualitatively different equation 
of state, namely, that of a zero-temperature relativistic Fermi gas. Note that we continue to focus on a one-component, self-gravitating 
system, so the same fermions that provide the source of gravity are also responsible for the pressure support that prevents the gravitational 
collapse of the star. In this way, one preserves the simplicity and internal consistency of the polytropic framework. However, unlike 
polytropes, the relativistic Fermi gas is not characterized by a single power-law relation between pressure and density, but rather 
by a continuous change in the central pressure of the star. By varying the central Fermi momentum (a proxy for the central density), 
one generates a continuous family of stellar configurations within a unique equation of state, ranging from the nonrelativistic to the 
ultrarelativistic regime.

The extension to Fermi gases is particularly useful for two reasons. First, it provides a non-polytropic test of the enthalpy formulation and 
therefore demonstrates that the method is not tied to the special scaling properties of Lane--Emden-type models\,\cite{Lane1870,Emden1907}. 
Second, the relativistic Fermi gas interpolates naturally between the familiar nonrelativistic and ultrarelativistic polytropic limits. It thus offers 
a controlled setting in which the stellar density can evolve without changing the underlying equation of state. This makes the model a useful 
benchmark for assessing the robustness of the Newtonian I-Love-Q relations beyond the class of polytropes, while retaining a framework 
simple enough that the connection between microscopic physics and stellar structure remains transparent.

\begin{figure}[ht]
\centering
\bigskip
\includegraphics[width=0.99\linewidth]{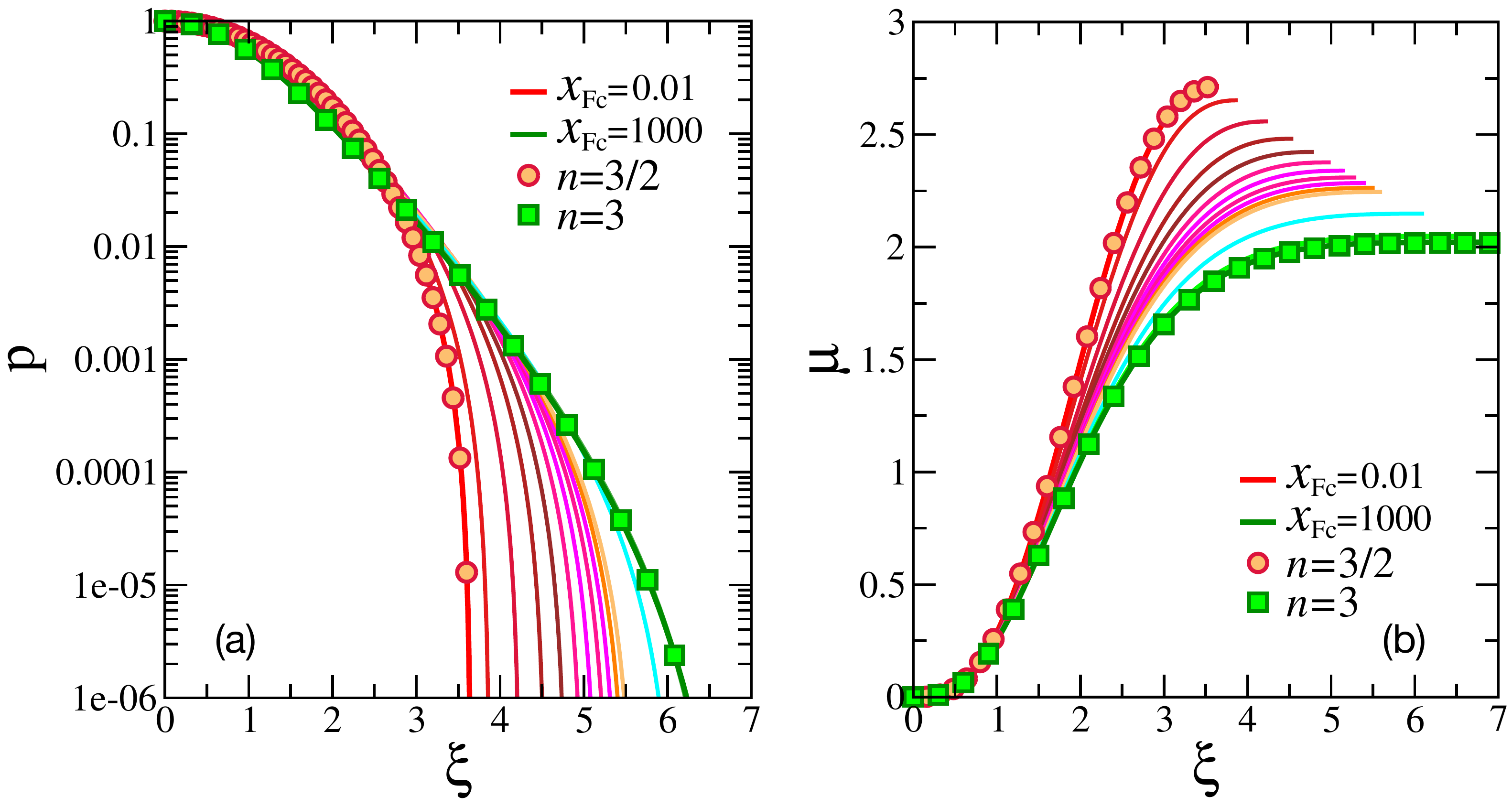}
\caption{(a) Dimensionless pressure profile and (b) mass profile of a zero-temperature relativistic Fermi gas as a function of the dimensionless
radial coordinate. The circles and squares denote the $n\!=\!3/2$ and $n\!=\!3$ polytropic limits, while the solid lines represent a sequence of
Fermi gas predictions that interpolates between these two limits.}
\label{Fig1}
\end{figure}

In Fig.\ref{Fig1} we display equilibrium configurations of a zero-temperature relativistic Fermi gas. Panel (a) displays the dimensionless 
pressure profile $p(\xi)$ as a function of the dimensionless radial coordinate $\xi\!=\!r/a$ while panel (b) shows the corresponding mass 
profile $\mu(\xi)$, both for a wide range of central Fermi momenta spanning the $x_{\rm Fc}\!=\!0.01$--$1000$ interval, where 
$x_{\rm F}\!\equiv\!p_{\rm F}/mc$ represents the dimensionless Fermi momentum. The circles and squares denote the $n\!=\!3/2$ and 
$n\!=\!3$ polytropic limits, respectively, so the figure nicely illustrates how the sequence interpolates smoothly between these two limits.

\begin{figure}[ht]
\centering
\bigskip
\includegraphics[width=0.8\linewidth]{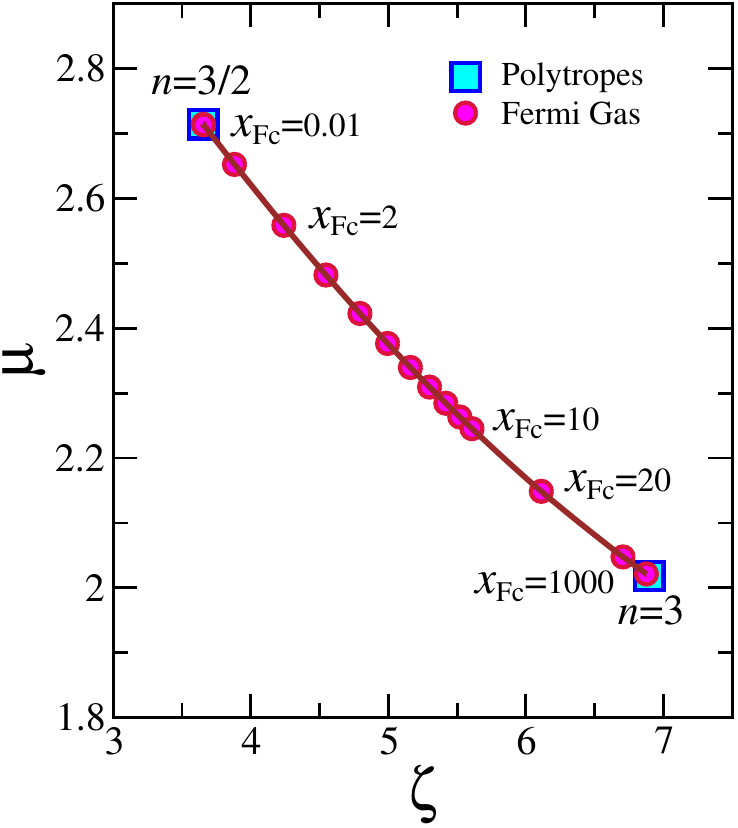}
\caption{Dimensionless mass-radius $\mu$--$\zeta$ relation as predicted by the same set of models used in Fig.\ref{Fig1}. 
	      The squares mark the two limiting polytropes, while the circles show the sequence of relativistic Fermi-gas models 
	      as the central Fermi momentum increases over the selected $x_{\rm Fc}\!=\!0.01$--$1000$ range.}
\label{Fig2}
\end{figure}

We now proceed to display in Fig.\ref{Fig2} the dimensionless mass-radius ($\mu$--$\zeta$) relation for the same set of models shown in
Fig.\ref{Fig1}. The squares denote the $n\!=\!3/2$ and $n\!=\!3$ polytropes, whereas the circles now represent the relativistic Fermi-gas 
sequence, with selected values of the central Fermi momentum $x_{\mathrm{Fc}}$ indicated for clarity. As $x_{\mathrm{Fc}}$ increases, 
the dimensionless radius $\zeta$ also increases while the dimensionless mass $\mu$ decreases, tracing a smooth curve between 
the two polytropic limits. 

These dimensionless trends should not be confused with the conventional mass-radius relation, because the length scale $a$ as well as 
the mass scale $4\pi\rho_c a^{3}$ both depend on the central Fermi momentum, which changes by several orders of magnitude over the 
adopted range. Nevertheless, by restoring units through $R\!=\!a\zeta$ and $M\!=\!4\pi\rho_c a^{3}\mu$, the common factor $a^3$ cancels 
in the ``mass-to-volume'' ratio:
\begin{equation}
 \frac{M}{R^3}=\frac{4\pi\rho_c\mu}{\zeta^3}.
 \label{MR3}
\end{equation}
Dividing the average density by the central density then provides a dimensionless measure of the evolving density profile:
\begin{equation}
 \frac{\langle\rho\rangle}{\rho_c} =\frac{3M}{4\pi\rho_c R^{3}} = 3\frac{\mu}{\zeta^{3}}.
 \label{muzeta3}
\end{equation}
The simultaneous decrease of $\mu$ and increase of $\zeta$ along the sequence imply a monotonic decrease of this ratio, from $0.167$ in 
the nonrelativistic $n\!=\!3/2$ polytropic limit to $0.0185$ in the ultrarelativistic $n\!=\!3$ limit, indicating the increasingly central concentration 
of the equilibrium configurations.
 
\begin{figure}[ht]
\centering
\bigskip
\includegraphics[width=0.95\linewidth]{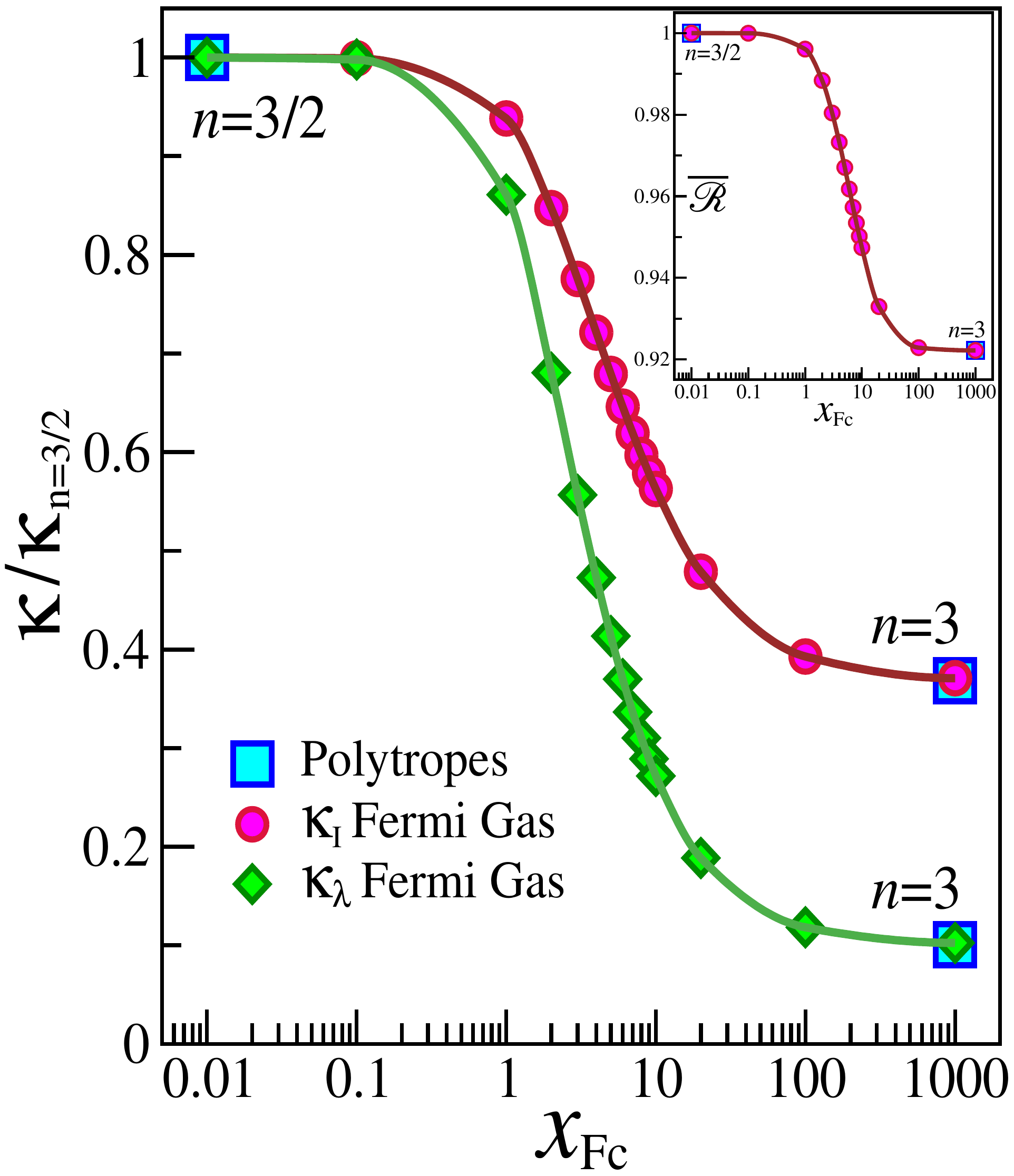}
\caption{Dimensionless structure factors $\kappa_I$ and $\kappa_\lambda$, normalized to the $n\!=\!3/2$ polytropic values, as a function 
of $x_{Fc}$. Circles and diamonds represent calculations with the relativistic Fermi-gas equation of state, while the squares denote
the polytropic limits. The inset displays the I-Love coefficient ${\mathcal R}\!\equiv\!{\mathcal R}_{I\lambda}$ also normalized to the 
$n\!=\!3/2$ polytropic value.} 
\label{Fig3}
\end{figure}

We finish this section by examining the Newtonian I-Love-Q relations through the ratios defined in Eqs.(\ref{RatiosKappa}).
These combinations eliminate the explicit dependence on the compactness parameter, which is sensitive to the equation of
state, leaving ratios of dimensionless structure coefficients. Motivated by the approximate universality established by Yagi and Yunes\,\cite{Yagi:2013bca,Yagi:2013awa}, one expects these remaining ratios to depend only weakly on the underlying 
stellar model. The present calculation tests this expectation along the Fermi-gas sequence. 

Notice, however, that within the Newtonian framework the two ratios are no longer independent but are related through the 
exact identity
\begin{equation}
 \frac{\bar{I}^{2}\bar{Q}}{\bar{\lambda}} =\frac{\kappa_I^{2}\kappa_Q}{\kappa_\lambda} =1.
 \label{ILQExact}
\end{equation}
Hence, it is sufficient to examine only the $\mathcal{R}_{I\lambda}$ coefficient defined in Eq.(\ref{RILambda}), given that the 
other ratio follows directly from the previous identity:
\begin{equation}
 \mathcal{R}_{Q\lambda} \equiv\frac{\kappa_Q}{\kappa_\lambda^{1/5}}
      = \frac{\kappa_\lambda^{4/5}}{\kappa_I^2} = \mathcal{R}_{I\lambda}^{-2}.
 \label{RQL}
\end{equation}

Figure\,\ref{Fig3} displays the dimensionless structure factors $\kappa_I$ (red circles) and $\kappa_\lambda$ (green diamonds),
each normalized to its corresponding $n\!=\!3/2$ polytropic value, as functions of the central Fermi momentum $x_{\rm Fc}$. Both 
coefficients interpolate smoothly between the two polytropic limits shown by the blue squares. In turn, the inset displays the I-Love 
coefficient $\mathcal{R}_{I\lambda}$, denoted in the figure simply by $\overline{\mathcal{R}}$. The inset illustrates that although 
the individual structure factors decrease substantially along the sequence, their I-Love combination decreases by only about $8\%$. 
This result indicates that much of their sensitivity to the changing density profile cancels in this combination, although a residual 
dependence remains as the configurations become increasingly closer to the ultrarelativistic $n\!=\!3$ limit. Given that the 
sequence is generated with a single Fermi-gas equation of state, this result demonstrates the relative insensitivity of the I-Love 
coefficient to structural changes along this sequence. It complements the comparison among polytropes in Table\,\ref{Table1} and 
the broader analysis of Ref.\cite{Yip:2017}. 

The exact identity in Eq.(\ref{ILQExact}) should be distinguished from the approximate constancy of the individual ratios. In the
Newtonian framework, slow uniform rotation and a weak tidal field probe the same linear quadrupolar response, with their driving 
sources simply related by $\omega^{2}/3\!\leftrightarrow\!\mathcal{E}/2$. The dependence on the internal 
structure therefore cancels exactly in $\bar{I}^{2}\bar{Q}/\bar{\lambda}$, irrespective of the underlying equation of state. Thus, 
the EOS independence of this identity follows from the common $l\!=\!2$ response, whereas the approximate constancy of the 
pairwise I-Love-Q coefficients involves an incomplete cancellation of the sensitivity to stellar structure. Although certain aspects 
of this cancellation and its breakdown have been elucidated in Ref.\cite{Yip:2017}, a general physical explanation of the residual 
EOS dependence remains incomplete.

\section{Conclusions}
\label{Sec:Conclusions}

The main goal of this paper was to provide a pedagogical introduction to the
I-Love-Q relations within the context of Newtonian gravity, emphasizing the
physical connections among stellar equilibrium, rotational deformation, and
tidal response. Although a quantitative description of these properties for 
neutron stars requires general relativity, the Newtonian regime greatly simplifies 
both the formalism and its physical interpretation. Our purpose has been to 
organize established results within a unified framework that distinguishes exact
consequences of the stellar response equations from approximate insensitivity
to the equation of state.

The central ingredient of this framework is the specific enthalpy $h$. In contrast
to the traditional treatment which enforces hydrostatic equilibrium through
Euler's equation, the introduction of $h$ converts Euler's equation into a first 
integral of hydrostatic equilibrium. Together with Poisson's equation and an
EOS in the form $\rho=\rho(h)$, these relations describe both the static spherical
configuration and its deformation under both rotation and tides. 

Within this enthalpy-based formulation, slow uniform rotation and a weak static
quadrupolar tidal field excite the same linear quadrupolar response. Their
respective external potentials have identical spatial dependence in the
quadrupolar sector and differ only in amplitude. Adopting dimensionless
normalizations, this equivalence yields the exact Newtonian identity
$\bar{I}^{\,2}\bar{Q}/\bar{\lambda}\equiv1$, irrespective of the EOS.
Importantly, this exact constraint does not require the pairwise I-Love-Q relations 
to be identical for different equations of state. Rather, their approximate universality 
involves an additional cancellation depending on the internal stellar structure.

We illustrated these results with both polytropic and relativistic Fermi-gas EOSs. 
The analytic $n\!=\!1$ polytrope provided a benchmark for the formalism and
its numerical implementation, while representative polytropes displayed both the
weak EOS dependence for relatively stiff equations of state and its deterioration 
for softer configurations. The relativistic free Fermi gas extended the analysis beyond 
polytropic EOSs and connected smoothly the nonrelativistic $n\!=\!3/2$ and 
ultrarelativistic $n\!=\!3$ limits. Together, these examples demonstrated that the exact 
analytic identity persists even as the density profile and individual structure coefficients 
may change substantially.

A natural next step is to extend this work to the general relativistic domain, in
order to understand both the emergence and the physical underpinning of the
universal I-Love-Q relations. Further, it will be important to separate the role of 
relativistic effects from the cancellations responsible for approximate EOS 
independence already manifest in the Newtonian approach. Although the
relations are largely independent of the EOS, the breakdown of universality and
its limitations may carry important information on dense-matter microphysics, a
topic at the heart of nuclear science. The Newtonian framework developed here 
provides a transparent starting point for such a future investigation.


%

\begin{acknowledgments}\vspace{-10pt}
This material is based upon work supported by the U.S. Department of Energy Office of Science, 
Office of Nuclear Physics under Award Number DE-FG02-92ER40750.
\end{acknowledgments}

\end{document}